\documentclass[aps,prc,twocolumn]{revtex4}

\usepackage{epsfig}
\usepackage{dcolumn} 
\usepackage{amsfonts}
\usepackage{color}

\newcommand{\be}{\begin{equation}}
\newcommand{\ee}{\end{equation}}
\newcommand{\ba}{\begin{array}}
\newcommand{\ea}{\end{array}}
\newcommand{\bea}{\begin{eqnarray}}
\newcommand{\eea}{\end{eqnarray}}
\def\hf{\textstyle{1\over2}}
\def\half{{1\over2}}

\begin{document}


\title{Quasi dynamical symmetry in an interacting
boson model phase transition}


\author{D.J.~Rowe}
\affiliation{Department of Physics, University of Toronto\\
Toronto, Ontario M5S 1A7, Canada}
\date{Feb.\ 5, 2004}

\begin{abstract} The oft-observed persistence of symmetry properties in
the face of strong symmetry-breaking interactions is examined in the 
 SO(5)-invariant interacting boson model.  This model
exhibits a transition between two phases associated with 
U(5) and O(6) symmetries, respectively, as the value of a control parameter
$\alpha$ progresses from 0 to 1.  The remarkable fact is that, for
intermediate values of $\alpha$ the model states exhibit the
characteristics of its closest symmetry limit for all but a relatively
narrow transition region that becomes progressively narrower as the
particle number of the model increases.  This phenomenon is explained in
terms of quasi-dynamical symmetry.
\end{abstract}

\maketitle

There have been numerous recent studies of phase transitions in nuclear
models \cite{PTs,FI,CZ,Jolie,LG}. Being finite zero-temperature
many-particle systems, nuclei do not exhibit the sharp phase transitions
observed in condensed matter physics. Nevertheless, theoretical models
designed to describe nuclei for particular values of their parameters can
be extended to study their behavior as their parameters are varied, e.g.\
as the particle number goes to infinity. 

This letter focuses on a phase transition \cite{FI} in the
{\em interacting boson model\/} (IBM) of Arima and Iachello
\cite{AI}. The IBM comprises $N$ boson particles each of
which has two states: an $L=0$ ($s$-boson) state and an $L=2$
($d$-boson) state with five orientations.
The creation and annihilation operators for these bosons are denoted
$\{s^\dag , d^\dag_\nu; \nu = 0,\pm 1\pm 2\}$ and $\{s , d^\nu;
\nu = 0,\pm 1\pm 2\}$, respectively.  They satisfy the usual boson
commutation relations
\be [s,s^\dag] = 1\,,\quad [d^\mu,d^\dag_\nu] = \delta_{\mu\nu}\,,
[s,d^\dag_\nu] = [d^\nu, s^\dag] = 0 \,.
\ee
Thus, the Hilbert space of the IBM  carries an irreducible representation
(irrep) of the group U(6) and can be realized as a subspace of states of
$N$ quanta of a six-dimensional harmonic oscillator.
The group U(6)  has several subgroups and
different phases of the model can be associated with 
Hamiltonians that are invariant under different subgroups.
We consider the  Hamiltonian \cite{AI,FI}
 \be \hat H(\alpha) = (1-\alpha)\hat H_1 + \alpha \hat H_2 \,,
\label{eq:2}\ee with control parameter $\alpha$, where $\hat H_1$ is the
U(5)-invariant
$d$-boson number operator and $\hat H_2$ is the O(6)-invariant operator:
\be \hat H_1 = \hat n = \sum_\nu d^\dag_\nu d^\nu \,, \quad \hat H_2 =
\frac{1}{N}
\hat S_+ \hat S_- \,,\ee
where
\be \hat S_+ = \hf (d^\dag\cdot d^\dag - s^\dag s^\dag ) , \quad
\hat S_- = \hf (d\cdot d- ss) \,.
\ee 
The electric quadrupole moment operator is represented in the
model by
\be \hat Q_\nu = \frac{Z}{\sqrt{N}} (d^\dag_\nu s + s^\dag d_\nu)\,, \quad
\nu = 0, \pm1, \pm 2\,, \label{eq:Q}\ee
where $Z$ is a suitable norm factor (can be thought of as the charge).

\begin{figure}[ht]
\epsfig{file=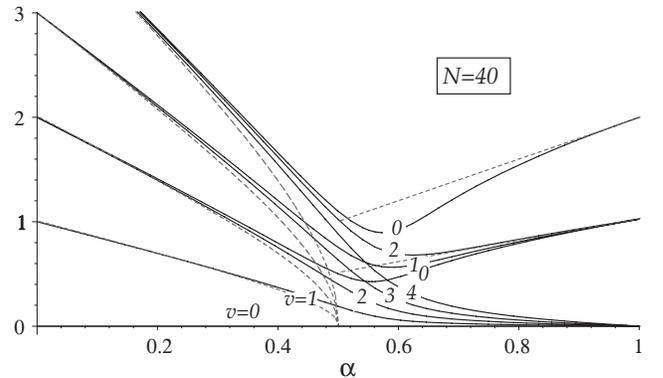, width=3.3 in}\hspace{0.75cm}
   \caption{Excitation energies of the Hamiltonian $\hat H(\alpha)$
plotted as a function of $\alpha$.    Precise, numerically computed,
energies are shown as continuous lines. The dotted lines are the results
of an RPA calculation (for $\alpha < 0.5$) and of another harmonic
aproximation (for $\alpha > 0.5$). \label{fig:1}}  
\end{figure}

\begin{figure}[ht]
   \epsfig{file=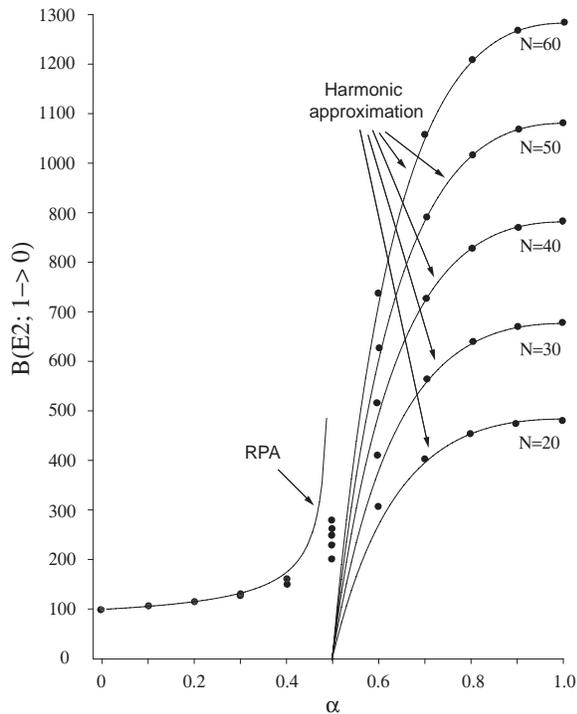, width=3.0 in}  
   \caption{$B$(E2) transition rates for decay of the first excited 
$v=1$ energy level to the ground state in the IBM for various values of $N$
expressed in units such that $B({\rm E}2; 1\to 0) = 100$ in the U(5) $(\alpha
=0)$ limit. The continuous lines for $\alpha \leq 0.5$ are for the RPA and
those for $\alpha \geq 0$ are for the harmonic approximation.
\label{fig:2}}   
\end{figure}

This letter shows the behavior of the energy-level spectrum
and the electric quadrupole transitions for the above model as $\alpha$ is
varied over the range $0\leq\alpha \leq 1$.
The  model is analytically solvable in its U(5) $(\alpha =0)$ and O(6)
$(\alpha=1)$ symmetry limits.
The numerically computed results given below for $0\leq\alpha\leq 1$
 are determined by use  of the simple
${\rm SU}(1,1)^s\times{\rm SU}(1,1)^d$ spectrum generating algebra with
basis elements
\begin{eqnarray}
&\begin{array}{c}
\hat S^s_+ = \textstyle\half s^\dagger s^\dagger , \quad
\hat S^s_- = \half ss , \\ \hat S^s_0 = \textstyle{1\over 4}( s^\dagger s
+ ss^\dagger )\,, 
\end{array}&\\
&\begin{array}{c}
\hat S^d_+ = \textstyle\half d^\dagger\cdot d^\dagger , \quad
\hat S^d_- = \half d\cdot d ,\\ \hat S^d_0 = \textstyle{1\over 4}(
d^\dagger\cdot d + d\cdot d^\dagger )\,. 
\end{array}&\end{eqnarray}
Energy levels, labelled by an SO(5) quantum number $v$, and E2
transition rates for decay of the first excited state of the model are
shown as a function of
$\alpha$ in figs.~\ref{fig:1}-\ref{fig:2}.

A superficial look at the results of figs.~\ref{fig:1} and \ref{fig:2}
would seem to suggest that the model holds onto its U(5) symmetry for
$0\leq\alpha\lesssim 0.3$ and to its O(6) symmetry for $0.8 \gtrsim \alpha
\leq 1$ and that there is a transition between the states of one symmetry
to the other in the intermediate region. It turns out this is an overly
simplistic view of what happens. Insight into the actual
evolution of the model states with increasing
$\alpha$ is given by approximate solutions which predict a phase
transition and do so with increasing accuracy as $N$ increases.
We consider the familiar Random Phase Approximation for $0\leq \alpha\leq
0.5$ and another harmonic approximation for $0.5 \leq\alpha\leq 1$.

For small values of $\alpha$, the RPA gives quasi-boson 
excitation operators 
\be D^\dag_\nu =(x\, d^\dag_\nu s - y\, s^\dag d_\nu )/\sqrt{N} \,,
\label{eq:6}\ee
with coefficients that satisfy an eigenvector equation of the
non-Hermitean form
\be \left( \begin{array}{cc}
A & B \\ -B & -A\end{array} \right) 
\left( \begin{array}{cc}x\\y\end{array}\right) = \varepsilon
\left( \begin{array}{cc}x\\y\end{array}\right) \,,
\ee
with submatrices given, e.g.\ in the
double-commutator equations of motion formalism \cite{EoM}, by
\bea &A_{\mu\nu} = {1\over N}\langle\phi|[s^\dagger d^\mu , [ H(\alpha),
d^\dagger_\nu s]] |\phi\rangle = \Big(1-{3\over
2}\alpha\Big)\delta_{\mu\nu} \,,\quad&\\ 
& B_{\mu\nu} = {-1\over
N}\langle\phi|[s^\dagger d^\mu , [ H(\alpha), s^\dagger d_\nu]]
|\phi\rangle =  -{1\over
2}\alpha\delta_{\mu\nu}\,,&
\eea
where  $|\phi\rangle$ is the uncorrelated
ground state given by the $s$-boson condensate  $|\phi\rangle =
(s^\dagger)^N |0\rangle / \sqrt{N!}$.
The RPA energy spectrum is shown in fig.~\ref{fig:1} as dotted lines for
$\alpha\leq0.5$.   It predicts a collapse of the excitation energies to
zero and, in fig.~\ref{fig:2}, a divergence of the E2 transition
rate for decay of the $v=1$ first excited state as $\alpha\to 0.5$.  Thus,
the RPA predicts a phase transition at $\alpha_{\rm
crit} = 0.5$ in accord with Thouless' Hartree-Fock stability condition
\cite{Thouless}.

\begin{figure}[ht]
   \epsfig{file=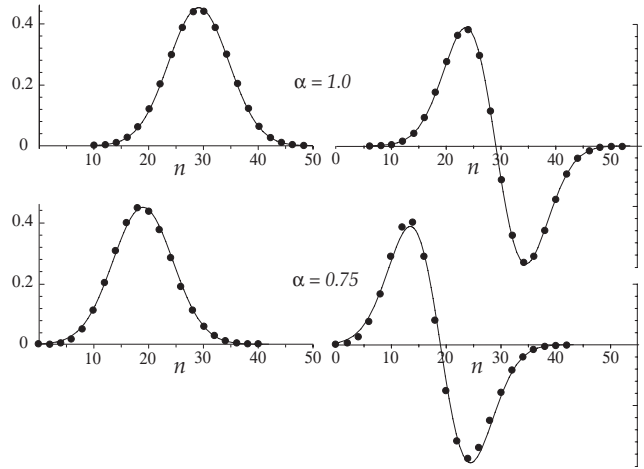, width=3.3 in}  
   \caption{Wave functions for the $N=60$ ground state and first excited
$v=0$ state for $\alpha = 1$ and $0.75$. The dots give the
coefficients of the states in a basis of $d$-boson number $n$.  The
continuous lines are those of the harmonic approximation.\label{fig:3}}  
\end{figure}

Details of the calculations will be given in an article to follow.  
It will then also be shown that there is another harmonic approximation
complementary to the RPA which gives the spectrum accurately in the
$\alpha \lesssim 1$ region.  The essential ingredients of this
approximation  are shown in fig.~\ref{fig:3} 
 which gives the expansion coefficients of the lowest and first excited
$v=0$ states for $\alpha = 1.0$ and 0.75 (cf.\ states labelled by $v$ in
fig.~\ref{fig:1}) in a basis $\{|nv\!=\!0\rangle\}$ which diagonalizes the
Hamiltonian $\hat H_1$ ($n$ is the $d$-boson number).  The smooth curves
through these numerically computed coefficients are simply harmonic
oscillator wave functions centred about $N/2$ for $\alpha =1$ and shifted
to smaller mean values of $n$, in the manner of
coherent state wave functions, as $\alpha$ is decreased.  The energy
levels and E2 transition rates predicted by this approximation are shown
for $0.5 \leq\alpha\leq 1$ in figs.~\ref{fig:1}-\ref{fig:2}. They are
precise for $\alpha=1$ and become increasingly accurate for smaller
$\alpha$  with
increasing values of the boson number $N$.

An examination of the physical significance of the wave functions of
fig.~\ref{fig:3} reveals that they represent a nucleus with a large
mean quadrupole deformation at $\alpha =1$ that decreases  as
$\alpha$ falls and the nucleus moves towards a spherical equilibrium shape.
Such behavior is expected from the form of the classical
potential energy \cite{Diep}  corresponding to the Hamiltonian
(\ref{eq:2}). This potential, given as a function
\be  V_\alpha (\beta)= N\big[ \beta^2 + \alpha
(-2\beta^2+\beta^4) + 0(\frac{1}{N})\big] 
\label{eq:10}\ee
\begin{figure}[ht]
   \epsfig{file=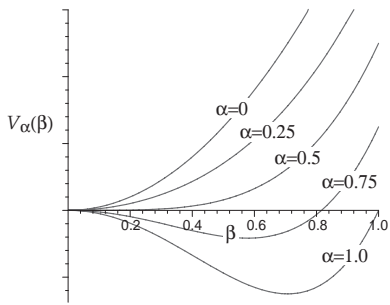, width=2.0 in}   
   \caption{The classical potential energy $V_\alpha(\beta)$  for different
values of $\alpha$. \label{fig:4}}  
\end{figure} 
of $\alpha$ and a classical quadrupole shape variable $\beta$, 
is shown for a few values of $\alpha$ in fig.~\ref{fig:4}.
The magnitude of the E2 (electric quadrupole) gamma-ray
transition rate shown in fig.~\ref{fig:2} is closely related to the mean
square quadrupole moment of the ground state and a good indicator of how
the mean value of the  quadrupole moment actually evolves with $\alpha$ in
the model.

Figure \ref{fig:3} reveals that the eigenfunctions of $\hat
H(\alpha)$ for $0.75 \lesssim \alpha \leq 1$ are close approximations
to harmonic oscillator (Glauber) coherent state wave functions. However,
as $\alpha$ falls below 0.75 (for $N=60$), the
wave functions reach the lower $n=0$ boundary and the harmonic
approximation begins to break down. For larger values of $N$, when the
potential $V_\alpha$ becomes deeper and the widths of the harmonic
oscillator wave functions narrower, the approximation holds for smaller
values of $\alpha$.  However, at $\alpha = 0.5$,  the centroid of the
wave functions of the harmonic approximation lie precisely
at $n=0$; thus the model breaks down as $\alpha\to 0.5$ for all values of
$N$. This is what one would expect for the
quantum states of a model with potential energy given by
eqn.~(\ref{eq:10}).

Similar insights can be gained for small $\alpha$ from the RPA.
For large values of $N$, in which the RPA is equivalent to the
Bogolyubov approximation of replacing the $s$-boson operators $s^\dag$
and $s$ by the c-number $\sqrt{N}$, the RPA transformation of
eqn.~(\ref{eq:6}) is seen as an SU(1,1) transformation
\be d^\dag_\nu \to D^\dag_\nu = x\, d^\dag_\nu  - y\, d_\nu \,.
\ee 
The corresponding transformation of the ground state  (to a
$D$-boson vacuum) is then a transformation to a dilated
(anti-squeezed) coherent state in which the equilibrium quadrupole shape
remains spherical but the width of its (harmonic oscillator Gaussian) wave
function, along with the magnitude of the quadrupole shape fluctuations
grow with increasing $\alpha$.
As $\alpha$ approaches the critical value 0.5 from below, the spherical
equibrium shape, indicated by the potential shown in fig.~\ref{fig:4},
becomes unstable and the width of the $D$-boson vacuum wave function
diverges.

The precision of the RPA and the other harmonic approximation, for
low-energy model states outside of a transition region that becomes
increasingly  narrow as $N\to\infty$, suggests a useful definition of the
concept of {\em phase\/} in such situations.

\medskip{\em Definition (phase):}
If SGC is a subgroup chain 
\be G_1 \supset G_2 \subset \cdots
\ee
of transformations of a system, 
then a subset of states of the system is said to be in an
SGC-phase if the properties of the subset are indistinguishable (to within
a specified accuracy) from a corresponding subset of states of a model
whose eigenstates belong to irreps of the subgroups in the chain and
whose observables are infinitesimal generators of $G_1$. 
\medskip

The above results show that there is
domain of $\alpha$ extending from 0 to some upper limit below 0.5 for
which a subset of low-energy states are described accurately by the RPA.
The RPA formalism shows that these states are equivalently described by a
U(5)-invariant model Hamiltonian 
\be  \hat H_{\rm RPA} = \varepsilon \sum_\nu D^\dag_\nu D^\nu + {\rm
constant}\,,\ee  with electric quadrupole operator
\be   \hat Q^{\rm RPA}_\nu = \frac{Z}{\sqrt{N}} (D^\dag_\nu s + s^\dag
D_\nu)\,,
\quad
\nu = 0, \pm1, \pm 2\,. \label{eq:QRPA}\ee
Thus, according to the definition, the subspace of states that are
described to within the required limits of accuracy by the RPA, for a given
value of $\alpha$,  belong to a ${\rm U(6)} \supset {\rm U(5)}\supset{\rm
SO(5)}$ phase.

It is similarly shown that the subset  of states
whose energies cluster around those of the harmonic approximation for
$\alpha
\lesssim 1$ span an O(6) irrep (for some choice of O(6)). Moreover, to the
extent that the harmonic approximation gives an accurate description of
a subset of states in this region, the energy levels and E2 transitions
between these states  are accurately described by a model whose states are
labelled by the quantum numbers of a ${\rm U(6)} \supset{\rm
O(6)}\supset{\rm SO(5)}$ subgroup chain (the small spread of energies of a
cluster is readily accommodated by including a term proportional to the
${\rm SO(5)}\subset{\rm O(6)}$ Casimir invariant in model
Hamiltonian). Thus, by the definition, the states that are accurately
described by the harmonic approximation,  belong to a 
${\rm U(6)} \supset{\rm O(6)}\supset{\rm SO(5)}$ phase.

The above analysis of the nature of the solutions of the IBM Hamiltonian
(\ref{eq:2}) in terms of the RPA shows why, in spite of the fact that
the U(5) symmetry of $\hat H(0)=\hat H_1$ is
quickly broken by the $\alpha\hat H_2$ term in $\hat H(\alpha)$ for $\alpha
>0$,  the results look as though the U(5) symmetry is preserved.
Similarly, the harmonic approximation of $0.5\gtrsim \alpha \leq 1$ shows
why the apparent O(6) symmetry is also retained over a considerably
larger region of $\alpha$ than expected.

An apparent persistence of a symmetry when numerical calculations show
there to be strong mixing of the irreps of the symmetry group in
question has been observed many times.
It has been expressed in terms of what has been dubbed {\em
quasi-dynamical symmetry} \cite{QDS1}; cf.\ ref.~\cite{QDS2} for a review.

A sense of what quasi-dynamical symmetry means is obtained by recalling
that  different but equivalent irreps of a group can be combined
coherently to give   new (equivalent) irreps.
A simple example would be the coherent mixing of the basis states       
$\{ |a LM\rangle\}$ and $\{|bLM\rangle\}$ of two
irreps of SO(3) of angular momentum L to form new basis states
\be |cLM\rangle = C_a |aLM\rangle + C_b |bLM\rangle \ee
for an equivalent mixed irrep of the same $L$;
coherent  mixing means that the coefficients $C_a$ and $C_b$ are
independent of $M$.
Quasi-dynamical symmetry arises because many groups have distinct irreps
that are similar and difficult to distinguish, especially by consideration
of a subset of their states. Thus, for example, it is possible to represent
many states of  a free particle with  wave functions of good linear
momentum even though the real states are wave packets comprising coherent
mixtures of plane wave functions of similar momentum.

It is not surprising then to find that subsets of states of a given system
can be described by models with dynamical symmetries that are only
quasi-dynamical symmetries of the sytem being described.
Because a model can at best describe a subset of the states of a
real physical system to within limited accuracy, it could not be
otherwise.
However, an explicit recognition of the possible quasi-dynamical
symmetries  of physical systems is invaluable for
interpreting the implications of successful models and in designing
successful approximations. In this note, I have
explicitly determined realizations of the quasi-dynamical symmetries of
two phases of the IBM and thereby obtained an explanation of why these
symmetries appear to persist in spite of the known presence of strong
symmetry mixing interactions.  Simply stated, the mixed states of the
original dynamical group become  the unmixed states of a quasi-dynamical
group. 

The above perspective leads one to think of the evolution of the
low-energy states of the IBM as they progress though a phase
transition in terms of the evolution of the
quasi-dynamical group.  In approximate pictorial terms, the effect of the
symmetry-breaking interaction is primarily to distort the quasi-dynamical
symmetry, rather than break it until a critical point is reached  at which
it can be distorted no more.  At this point  the symmetry really starts to
break up; the system enters the transition region and, as it
emerges on the other side, a new quasi-dynamical symmetry associated with
another other phase  begins to develop.

An interesting characteristic of the above results  is that for reasonably
large values of the boson number $N$, the  low-energy states
of the model can be assigned to one of three domains: one for which the 
${\rm U(6)} \supset {\rm U(5)} \supset {\rm SO(5)}$ phase is appropriate,
one for which the
${\rm U(6)} \supset {\rm O(6)} \supset {\rm SO(5)}$ phase is appropriate,
and an intermediate transition domain which shrinks with increasing $N$.
A recent suggestion of critical point symmetries \cite{FI} which apply in
the middle of the transition region is therefore of considerable interest;
for example, it suggests that it might be meaningful to think of an
intermediate phase separating the other two.
This suggests that an interesting sequel to the present investigation
would be an exploration of the ways the properties of the system in the
transition domain  evolve as a function of the boson
number $N$.

\begin{acknowledgments}
The author wishes to thank J.L.\ Wood for helpful discussion.
\end{acknowledgments}

\vfill

\end{document}